\documentclass[aps,prl,twocolumn,preprintnumbers,superscriptaddress,10pt,longbibliography]{revtex4-2}
\usepackage{color}
\usepackage[utf8]{inputenc}
\usepackage{hyperref}
\usepackage{amsmath}
\usepackage{graphicx}

\usepackage{algorithm}
\usepackage{algorithmic}

\newcommand{\eq}{\begin{equation}}
\newcommand{\eqx}{\end{equation}}
\newcommand{\eqn}{\begin{eqnarray}}
\newcommand{\eqnx}{\end{eqnarray}}
\newcommand{\f}[2]{\frac{#1}{#2}}

\newcommand{\al}{\alpha}

\newcommand{\cor}[1]{\left\langle{#1}\right\rangle}

\begin{document}


\title{On  the linear scaling of entropy vs. energy in human brain activity, the Hagedorn temperature and the Zipf law }



\author{Dante R. Chialvo}
\email{dchialvo@gmail.com}
\affiliation{Instituto de Ciencias F\'isicas (ICIFI-CONICET), Center for Complex Systems and Brain Sciences (CEMSC3), Escuela de Ciencia y Tecnolog\'ia, Universidad Nacional de Gral. San Mart\'in, Campus Miguelete, 25 de Mayo y Francia, 1650 San Mart\'in, Buenos Aires, Argentina}
\affiliation{Consejo Nacional de Investigaciones Cient\'{\i}fcas y Tecnol\'ogicas (CONICET),  Godoy Cruz 2290, 1425 Buenos Aires, Argentina}
\author{Romuald A. Janik}
\email{romuald.janik@gmail.com}
\affiliation{Institute of Theoretical Physics and Mark Kac Center for Complex Systems Research, Jagiellonian University,  ul. {\L}ojasiewicza 11, 30-348 Krak{\'o}w, Poland}

\begin{abstract}
 It is well established that the brain spontaneously traverses through a very large number of states. Nevertheless, despite its relevance to understanding brain function, a formal description of this phenomenon is still lacking. To this end, we introduce a machine learning based method allowing for the determination of the probabilities of all possible states at a given coarse-graining, as well as the density of states, from which all the thermodynamics can be derived. This is a challenge not unique to the brain, since similar problems are at the heart of the statistical mechanics of complex systems. This paper provides a rigorous demonstration of the linear scaling of the entropies and energies of the brain states, a behaviour first conjectured by Hagedorn to be typical at the limiting temperature in which ordinary matter disintegrates into quark matter. Equivalently, this establishes the thermodynamics origin of the Zipf law scaling underlying the appearance of a wide range of brain states. Based on our estimation of the density of states for large scale functional magnetic resonance imaging (fMRI) human brain recordings, we observe that the brain operates asymptotically at the Hagedorn temperature. The presented approach is not only relevant to brain function but should be applicable for a wide variety of complex systems.
\end{abstract}
\maketitle


The proposition that the brain is poised near a critical point of a phase transition suggests that properties universally observed near criticality can be advantageous for brain function. Those include, among the most important, the very large number of metastable states present at criticality,  which provides for large memory capacity, also the large susceptibility which may endow the brain with an exquisite sensitivity and at the same time an unlimited dynamic range to process sensory inputs. Finally a rough energy landscape can provide high flexibility for eventual swift changes.   Since the hypothesis  was put forward by Bak and colleagues \cite{bak1}, considerable experimental support for the main idea has been accumulated, as well as some constructive controversy over technical aspects (see~\cite{BIALEK1,Chialvo2010,review1,review2, plenz,haimo,taglia1,camargo}). In this paper we focus on a model-independent feature of criticality, specifically in the space of states that the human brain explores during  the so called resting conditions (i.e. un-purposeful behaviour) using functional magnetic resonance imaging (fMRI) recordings.  According to current understanding, such space of states supposedly represents the brain (microscopic) dynamical repertoire available for perception, cognition and behaviour, hence its importance for understanding brain function in health and disease.
\begin{figure}[t!]
\includegraphics[width=0.49\textwidth]{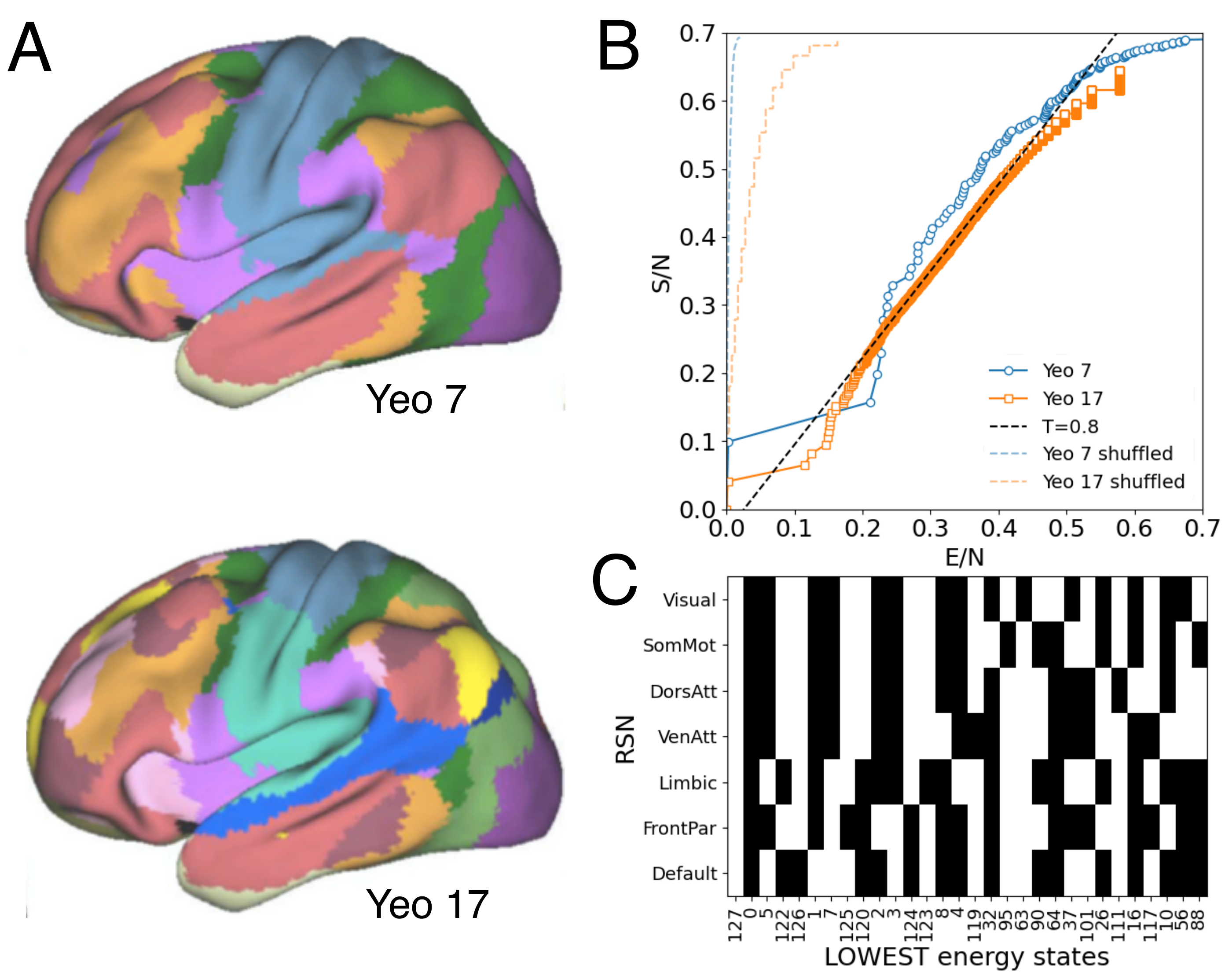}
\caption{Entropy-energy dependence (panel B) for very coarse grain fMRI brain data (obtained from the 7 or 17 Yeo's brain parcellations \cite{Yeo}, snaphots shown in panel A) computed by state counting of long concatenated recordings. Compare with the null hipothesis plotted with dashed lines corresponding to results gathered from randomly shuffling the data. Panel C shows the 30 lowest (of $2^{7}$) energy states (with white/black denoting active/inactive) computed for the Yeo's 7 parcellation. 
\label{fig.rsn}}
\end{figure}

In contrast to standard statistical systems like the Ising model, where the concrete discrete space of states is provided from the outset by its definition, the space of states of the brain is an approximate concept which is relative to the employed level of analysis and the problem of interest.
For the analysis in the present paper, we binarize resting-state fMRI recordings spatially parcellated into various regions of interest (ROIs), so that a state is a binary configuration being +1 for ROI's which are predominantly active at that instance of time and -1 otherwise (see below for details). We analyze these spaces of binary states extracted at various levels of granularity from 7 or 17 Resting State Networks up to fine parcellations into 1000 regions. We use resting state fMRI data of 100 subjects released by the Human Connectome Project~\cite{HCP} which once concatenated results in a massive dataset of 456000 states occurring during resting-state brain activity.

A key finding of the paper is the most rigorous demonstration at present that the resting state brain activity obeys a Zipf law, well known from linguistics, i.e. the frequencies of states are approximately a power function of their rank
\eq
frequency \sim \f{1}{rank^s}
\label{e.zipf}
\eqx
This is equivalent to a linear entropy-energy $S(E)$ behaviour (see below) and clearly manifests itself in the resting-state data as in Fig.~\ref{fig.rsn}B for Resting State Networks and in Figs~\ref{fig.schaefer}A-\ref{fig.PRG}A for finer parcellations.

In view of leveraging concepts from statistical physics and using the binary character of the fMRI brain state constituents (which can be thought of as ``spins''), it is illuminating to introduce an effective Hamiltonian such that the discretized fMRI signals can be treated as samples from a Boltzmann distribution at a fixed reference/operating temperature $T_{ref}=1$. 

The second key contribution of the present paper is to introduce a Machine Learning algorithm extending~\cite{ENTROPYML} for estimating \emph{both} the effective hamiltonian and the density of states (equivalently $S(E)$). This allowed us to extend the results for Resting State Networks (Fig.~\ref{fig.rsn}B) to much more fine-grained and thus high dimensional signals with spaces of states numbering up to $2^{1000}$ possible elements (Figs~\ref{fig.schaefer}A-\ref{fig.PRG}A). 
The Machine Learning method introduced here, should in fact be of a very general applicability. 

The above statistical physics interpretation and the observation of a \emph{linear} $S(E)$ curve (and thus Zipf law scaling), leads to an exponential density of states
\eq
\rho(E) = e^{S(E)} \sim e^{\f{1}{T_H} E}
\label{e.hagedorn}
\eqx
characteristic of so-called Hagedorn behaviour with Hagedorn temperature $T_H$. 
It is very intriguing that the Hagedorn behaviour, appearing in various contexts in high-energy physics, is also clearly realized in the context of the brain.

The third key contribution of the present paper is that for fine-grained parcellations, the observed Hagedorn temperature approaches the reference/operating temperature $T_{ref}=1$ of the effective Boltzmann ensemble (see Fig.~\ref{fig.PRG}B). This finding indicates that the brain operates asymptotically at the Hagedorn temperature.

The plan of the paper is as follows. First, we review the interrelations of the density of states and entropy-energy curves with the Zipf law and Hagedorn temperature. Second, we introduce an extension of the Machine Learning (ML) method of~\cite{ENTROPYML} so that it can extract both the energy and entropy from extremely high-dimensional binary data and verify its reliability on solvable examples of 1D and 2D Ising models. Third, we apply the ML method to the resting-state fMRI data and present our results. We close the paper with a summary and discussion.

\section{Interrelations of density of states, entropy-energy curves and Hagedorn temperature}

A key characteristic of the system's space of states is the \emph{density of states} as a function of energy, which determines the thermodynamic behaviour of the system at any temperature, including the existence of criticality.
Typically, when studying systems in the thermodynamic limit of infinite size, the system explores only a relatively small portion of the space of states (largest in the critical case). When we deal with a finite but large ``mesoscopic'' system, however, the repertoire of explored states may be wide enough to encompass information on the behaviour of the system for a range of temperatures. In particular, it could detect signatures of critical behaviour even away from the system's operating temperature.
In addition, it provides a much more detailed understanding of the statistical properties of the system,
hence its determination is of high interest.

The density of states is equivalently expressed in terms of the energy dependence of the (microcanonical) entropy as $\rho(E)=\exp S(E)$. 
In this language a second order phase transition is characterized by $S''(E_c)=0$ at some energy~$E_c$ in the thermodynamic limit. Of particular interest for the present paper is the case of a linear dependence 
\eq
S(E)\sim aE+b
\label{e.linear}
\eqx
On the one hand, as discussed in~\cite{BIALEK1, BIALEK2}, it is equivalent to Zipf law~\cite{ZIPF} -- a power-law dependence of the probability of a state on its rank~\eqref{e.zipf}.
The equivalence arises through an appropriate identification of energies and entropies for a given ensemble of binary data (see~\cite{BIALEK1, BIALEK2}), which we now briefly review.

The energies of states are identified with the logarithm of the state probabilities:
\eq
E(\{s_i\}) = -T_{ref}\,\log p(\{s_i\}) + c
\label{e.energy}
\eqx
where $T_{ref}=1$ is typically chosen for convenience and
$c$ is an arbitrary constant, which we set following~\cite{BIALEK2} so that the observed lowest energy state has vanishing energy.
The formula (\ref{e.energy}) effectively \emph{defines} a Hamiltonian such that the given data could be treated as arising from a Boltzmann distribution at the reference/operating temperature $T_{ref}=1$.
Eq.~\eqref{e.energy} is very informative as it allows to \emph{extract} the effective Hamiltonian of the system and hence all interactions directly from the probabilities of the particular brain states.
Of course, the utility of~Eq.\eqref{e.energy} depends on the ability of explicitly estimating the state probabilities. We would like to restrain ourselves from making any \emph{a-priori} assumptions about the form of the Hamiltonian (such as an Ising model with general pairwise couplings, which is the standard starting point for the maximum entropy method \cite{god,roudi}). We will instead use a very general Machine Learning algorithm (described later in the paper) for estimating the energies.

The remaining ingredient is the (microcanonical) entropy $S(E)$. Following the considerations of~\cite{BIALEK1}, we identify it with the logarithm of the rank of the corresponding state $\{s_i\}$ with energy~$E$:
\eq
S(E) \simeq \log rank(\{s_i\})
\label{e.SE}
\eqx

On the other hand, the linear entropy-energy relation~\eqref{e.linear} leads to an exponential density of states~\eqref{e.hagedorn}. Such a behaviour was first conjectured in the context of hadronic physics leading to the Hagedorn hypothesis of  a limiting  temperature ~\cite{HAGEDORN} (equal here to the inverse slope $T_H=1/a$), beyond which hadronic matter cannot exist. Since then, Hagedorn temperature appeared in various other contexts in high energy physics~\cite{HAGEDORNOTHER1,HAGEDORNOTHER2}. Cabbibo \& Parisi~\cite{parisi} found later that similar exponential behaviour can be present in any system which undergoes a second order phase transition, Note that  Ref.~\cite{BIALEK1} informally referred to this behaviour as \emph{very critical}, as then $S''(E)=0$ for a whole range of energies instead of just for a single point $E=E_c$.



\begin{figure}[t!]
\includegraphics[width=0.4\textwidth]{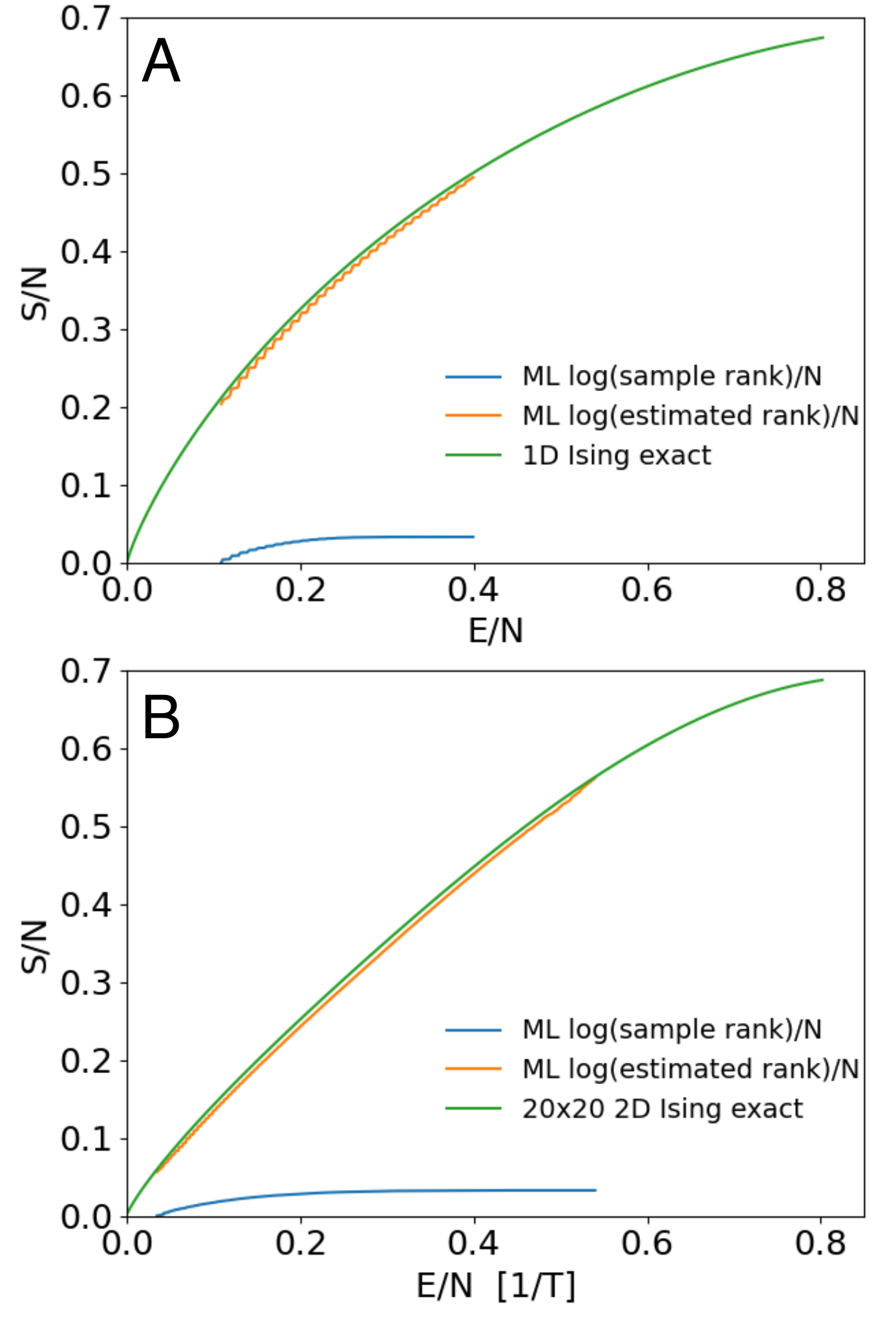}
\caption{Validation of the ML method results with the Ising model ground-truth expectations. Entropy-energy plots extracted using the ML  method for Monte Carlo simulations of the Ising model. Panel A shows the results for the non-critical 1D Ising ($T=1, N=400$)  and  panel B those for the critical 2D Ising model ($T=2.3, N=400$).
\label{fig.examples}}
\end{figure}



\section{Extracting energy and entropy. From Machine Learning to state probabilities and ranks}

To get insight on the system dynamics (in the brain and elsewhere), it is imperative to compute the probabilities from high dimensional sampling.   Of course,  any naive attempt of doing it from occurrence counts is invalidated by  the exponential increase in the number of states, resulting in the vast majority of states appearing only once in the sample data. 
In order to overcome this difficulty, we adapt a recently developed method of using machine learning classifiers to compute entropy from Monte Carlo samples valid even when all samples are distinct~\cite{ENTROPYML}. As the machine learning classifiers used to compute the overall Shannon entropy directly estimate the probability of each state, they are a perfect tool for determining~Eq.\eqref{e.energy}.

We now review briefly the approach of~\cite{ENTROPYML} of computing Shannon entropy of high dimensional signals and show how it can be adapted to evaluate the probabilities of individual states, as well as extended for estimating the ranks of states necessary for computing~\eqref{e.SE}.

The key difficulty in computing the Shannon entropy
\eq
S_{Shannon} = \cor{- \log p} = -\sum_\al p_\al \log p_\al
\eqx
is reliably estimating the $\log p_\al$ of states $\al \equiv \{s_i\}_{i=1}^N$ appearing in the data, where $N$ is the dimensionality and $s_i=\pm 1$. In the high dimensional case 
we use machine learning tools to analyze the \emph{internal structure} of the states instead of treating each state as an opaque entity
when using occurrence counts.
To this end, we employ the standard decomposition of multivariate probability
\eq
p(s_1, \ldots, s_N) = p(s_1)p(s_2|s_1)p(s_3|s_1, s_2) \ldots
\eqx
Taking the logarithm, we see that we have to evaluate the conditional probabilities like $\log p(s_3|s_1, s_2)$. But this is just what virtually any machine learning classification algorithm provides when predicting the spin $s_3$ based on $s_1$ and $s_2$.
Training $N$ such classifiers leads to a prescription for $\log p(s_1,\ldots, s_N)$, which can be then evaluated for any state of interest to obtain its energy according to~\eqref{e.energy}.
Evaluating the average of $\log p(s_1,\ldots, s_N)$ over the data sample gives the estimate of Shannon entropy from~\cite{ENTROPYML}. Care must be taken to always evaluate predictions on held-out out-of-sample data. See \textit{Supplementary Information} for details.

For the purposes of the current paper, we need a new ingredient which was not present in the considerations of~\cite{ENTROPYML}, namely evaluating the rank of particular states $rank(\{s_i\})$ appearing in~\eqref{e.SE},
yielding simultaneously an estimate of the density of states at a given energy level.
The main difficulty is the fact that the provided data sample contains usually only a very small subset of all possible states, while the rank should be computed in the space of all states, including many unseen ones.

In order to quantify the unseen states, we will assume that each state from the data sample of a given energy~$E_\al$ is a representative of some multiplicity of states $n^{mult}_\al$ of similar energy in the whole space of states. 
We estimate these multiplicities by requiring the consistency of computing Shannon entropy from the average of the predicted $\log p_\al$ over the provided data sample, as was done in~\cite{ENTROPYML}, with the one computed purely from the predicted $p_\al$'s and the multiplicities.
We thus have
\eq
-\sum_{\al\in data} p^{empirical}_\al \log p_\al  \simeq -\sum_{\al\in data} p_\al n^{mult}_\al \log p_\al
\label{e.twoentropies}
\eqx
where the summation goes over the distinct states present in the data sample and $p^{empirical}_\al$ is based on the occurrence count of state $\al$. The consistency of the two formulas leads to the expression
\eq
n^{mult}_\al = \f{p^{empirical}_\al}{p_\al}
\label{e.nmult}
\eqx
The estimate of the true rank is then obtained as the cumulative sum of the multiplicities leading to
\eq
S(E) \simeq \log \sum_{\substack{\al\\ E_\al\leq E}} n^{mult}_\al
\eqx

\begin{figure}[t!]
\includegraphics[width=0.4\textwidth]{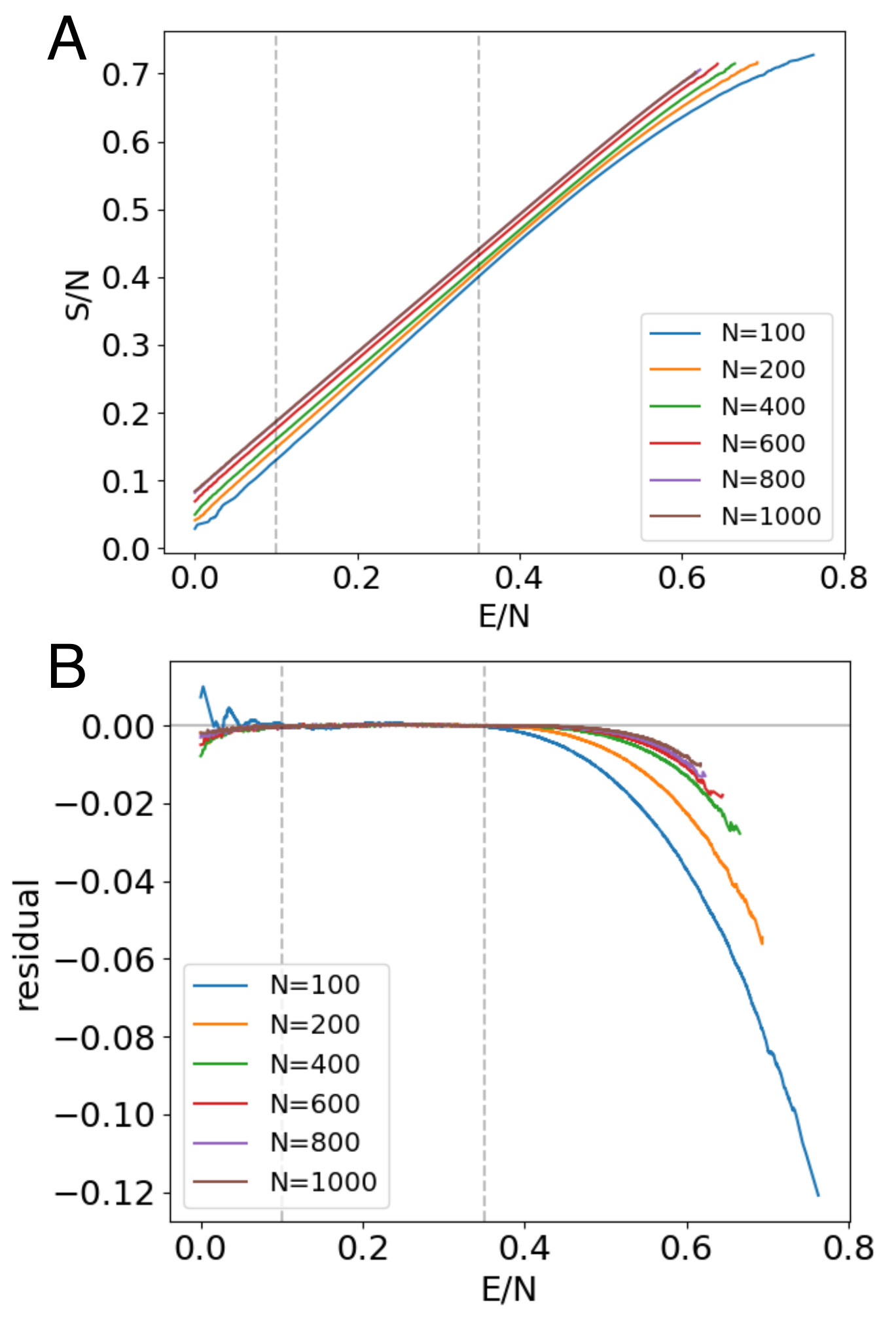}

\caption{Panel A: Entropy-energy plots for resting state fMRI data with Schaefer parcellations of varying sizes. Panel B: The residuals of the linear fits performed between the two dashed lines.  
\label{fig.schaefer}}
\end{figure}

\smallskip

\noindent {\textit {Validating the ML method with the Ising models.--}} 
In order to test the proposed ML method for obtaining the density of states, we applied it to the exactly solvable 1D and 2D Ising models with N=400 spins, which have similar dimensionality as the parcellated brain data and are respectively either always noncritical or close to criticality (evaluated here at $T=2.3$). 
The results are shown in  Fig.~\ref{fig.examples} (see more checks in the \textit{Supplementary Information}). Here, the arbitrary constant $c$ in~(\ref{e.energy}) was adjusted to agree with the conventions of the Ising models used for the ground-truth $S(E)$ curves. For the $N=400$ dimensionality, the vast majority of states are unobserved, so the estimate of the multiplicity~(\ref{e.nmult}) is crucial for obtaining the correct result in  Fig.~\ref{fig.examples}. Indeed, ignoring the multiplicities and using just the naive ranks of the observed states in the data, not only severely underestimates the entropy, but also leads to a different functional shape of the energy dependence (see the blue versus orange curves in Fig.~\ref{fig.examples}).

\section{Applying the ML method: Brain fMRI resting state data results}

We analyzed signals corresponding to fMRI timeseries which measure the level of activation of the brain (as a function of time) simultaneously at several thousands of contiguous locations.  We use resting state fMRI data of 100 subjects released by the Human Connectome Project~\cite{HCP} which once concatenated results in a total of 456000 timesteps (see \textit{Supplementary Information} for detailed information). The brain signals are usually coarse-grained (parcellated) at various scales~$N$ in accordance with functional or anatomical criteria.  Fig.~\ref{fig.rsn}A  illustrates two of the coarser \cite{Yeo} parcellations (also termed regions of interest ROI)  of $N=7$ and $N=17$, representing so-called Resting State Networks, from which mean time series for each parcel are extracted. Subsequently each signal is transformed into a dichotomous time series, by setting the samples above/below the median of each session to +1/-1 (active/inactive) so that the entropy of the resulting time series is maximized for each individual ROI.


For such heavily coarse-grained cases as the Resting State Networks, the probabilities of all states  can be estimated just by counting their number of occurrences. This is the case presented in Fig. \ref{fig.rsn} which shows that the brain entropy-energy plots, even at this low resolution, already exhibit  a region of linear scaling. The slope is above 1, indicating that the corresponding critical (Hagedorn) temperature is \emph{lower} than $T_{ref}=1$. 
Note that this is not in contradiction with the concept of the Hagedorn temperature as a limiting temperature, as we are dealing here with a finite system.
This result directly shows the appearance of Zipf law/linear scaling for a subset of states without recourse to any nontrivial machine learning based methods.
Decorrelating the time series by shuffling the individual RSN signals, clearly completely destroys this behaviour as shown by the dashed curves in Fig.~\ref{fig.rsn}B.
Notice, also, that the two lowest energy vectors (panel C) correspond, as expected, to different degrees of global synchronous activation or silence.  Despite of its relatively large coarse-graining, the analysis is able to reveals a gradient in the energy of activation patterns from  the (more associative) limbic, frontoparietal and default mode areas to those corresponding to primary (somatosensory, visual and attention) areas.

As commented above, for parcellations with small $N$, we can estimate the state probabilities just from the occurrence counts and directly compute the corresponding ranks. 
In fact, the case of $N=17$ shown  in Fig. \ref{fig.rsn} is at the borderline of the applicability of the direct methods, as we found that only $\sim 43\%$ of the possible states occur in the data. 


Now we proceed to the main challenge of the paper, which is to analyze states for much finer parcellations. First, we consider the family of Schaefer parcellations constructed in~\cite{SCHAEFER}, as they come in variants ranging from sizes $N=100$ to $N=1000$ ROIs. 
Such parcellated time-series contain only a very small fraction of all the possible states and even for the smallest $N=100$ case, we found that $\sim 98\%$ of the states in the data appear only once. Thus we have to employ the Machine Learning method described earlier in the paper. 
We use a state-of-the-art nonlinear gradient-boosted-tree classifier \texttt{xgboost}~\cite{xgboost} with settings as in~\cite{ENTROPYML} (see \textit{Supplementary Information} for details).

The exploration of the Schaefer parcellations for a variety of sizes  (see Fig.~\ref{fig.schaefer}A) reveals a linear behaviour whose range increases with $N$ (see Fig.~\ref{fig.schaefer}B), confirming the observation made for the $N=17$ Resting State Networks shown in Fig.\ref{fig.rsn}B. 

\begin{figure}[!t]
\includegraphics[width=0.4\textwidth]{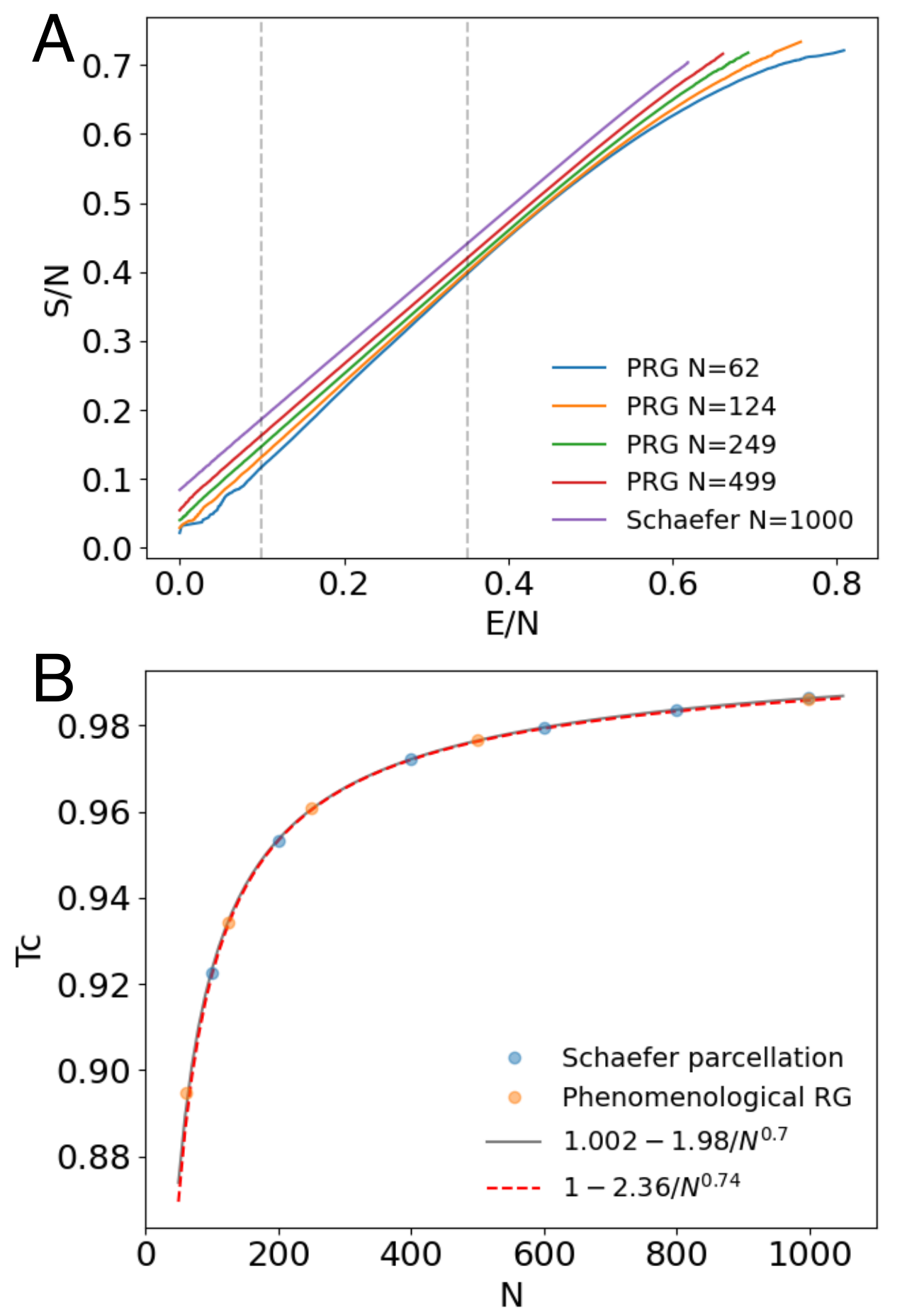}
\caption{Panel A: Entropy-energy plots for brain data successively coarse-grained using Phenomenological RG. Panel B shows the Hagedorn temperature extracted from the inverse slope of the linear fits as a function of parcellation granularity for both the PRG and the Schaefer parcelations (i.e., the data in Fig. ~\ref{fig.schaefer}A).  In both cases temperature scales with $N$ as a power law (the power law fits denoted in the legends were performed only using the Schaefer parcellations). 
\label{fig.PRG}}
\end{figure}

As a further test of the robustness of the results, we constructed a set of parcellations applying the Phenomenological Renormalization Group (PRG) approach of~\cite{PRG} resulting in subdivisions with $N=62, 124, 249$ and $499$ ROIs,  out of 998 ROIs of the $N=1000$ Schaefer parcellation. PRG parcellations differ from the Schaefer ones in that they are constructed following correlation hierarchies and often group together ROIs lying in different brain hemispheres. Remarkably, the critical Hagedorn temperatures extracted from the PRG subdivisions are very much in line with the expected values from the Schaefer parcellations (see Fig.~\ref{fig.PRG}B). The qualitative observation of the critical (Hagedorn) temperature approaching $T_{ref}=1$ with increasing $N$ is quantitatively confirmed by the power law fits 
\eq
T_H \sim T_{asymptotic} - \f{a}{N^b}
\label{e.powerlaw}
\eqx
shown in Fig.~\ref{fig.PRG}B. We find that the asymptotic Hagedorn temperature $T_{asymptotic}$ in the limit of maximally fine-grained parcellation (i.e. $N\to \infty$) is consistent with the reference/operating temperature of the Boltzmann distribution $T_{ref}=1$.
This indicates that, asymptotically at a fine-grained level, the brain effectively operates at the Hagedorn temperature with a very large repertoire of brain states satisfying Zipf law. This is one of the key results of the present paper.
The very clean power law behaviour observed in Fig.~\ref{fig.PRG}B and the specific value of the power law exponent $b\sim 0.7$ remain as an outstanding challenge for our understanding of brain dynamics.

A finding consistent with the appearance of the linear $S(E)$ regime close to $T_{ref}=1$, is the observation that the range of energies (scaled by $N$) present in the data sample is remarkably stable with increasing $N$. This is in marked contrast with the conventional picture in statistical physics where the range markedly decreases with~$N$, especially away from a critical point. Indeed, a linear dependence of $S(E) \sim E/T_H+const$ between $E_0$ and $E_1$ means that at $T=T_H$, all energies between $E_0$ and $E_1$ are equally likely to occur. This follows from rewriting the partition function as
\eq
Z(T) = \int dE \, e^{S(E) - E/T} \sim \int dE \, e^{E \left(\f{1}{T_H} - \f{1}{T}\right)}
\label{e.measure}
\eqx
Indeed for $T=T_H$ the measure in energy is flat. We address an interpretation of this result in the Discussion.

We performed a number of checks  (see \textit{Supplementary Information}) in order to confirm the robustness of the results, including de-correlating the data, testing the stability of the analysis w.r.t. the filtering of the fMRI data, and the degree of scale-invariance of successive parcellations. 

\section{Discussion}

In this work, we explore statistical aspects of the space of states of the human brain resting state activity and show the appearance of regimes of linear $S(E)$ dependence in the entropy-energy plane indicating proximity to a critical point (which is even \emph{very critical} in the terminology of~\cite{BIALEK1} i.e. $S''(E)=0$ for a range of energies instead of just a single point $S''(E_c)=0$). Moreover, we find that the brain operates asymptotically at the limiting Hagedorn temperature. The present analysis was made possible by a novel Machine Learning  based method for estimating the density of states from relatively high dimensional data (of the order of  hundreds of spins) which we developed as an extension of~\cite{ENTROPYML}, an  approach that should be, in fact, of a very general applicability.

Our considerations are very much model-independent, in particular we do not assume a reduction to pairwise interactions and we admit  in principle arbitrary higher order interactions.  Our method, however, does not require to specify these interactions explicitly, as they are encoded by the trained (nonlinear) machine learning classifiers.   
The only overall apparent assumption is that we treat the discretized fMRI signals as samples from a Boltzmann distribution at a fixed temperature.
Note, however, that this is really just a dictionary which translates the raw statistical quantities like state frequencies/probabilities and ranks of states in order of frequency into the language of statistical physics, thus allowing us to fruitfully leverage statistical physics concepts and intuitions in the context of the brain.  
The raw finding of a power-law probability-rank relation i.e. Zipf law is in fact completely independent of the use of the Boltzmann distribution interpretation of the data.

We find it very intriguing that Zipf law, first established in the context of linguistics i.e. a complex communication system, can also be clearly observed in the statistics of brain states appearing during resting-state activity of the human brain.
Of course, this type of behaviour may appear in various other contexts, like in the statistics of image patches in natural images~\cite{BIALEK2}. We should emphasize, however, that here we are analyzing resting-state fMRI data i.e. with the environmental input reduced to a minimum. Also the investigated brain states involve the activities of ROIs throughout the brain, not only sensory ones, so we can confidently rule out any external source of the observed behaviour. 

On the other hand, a derivation of Zipf law due to Mandelbrot~\cite{Mandelbrot} which rests on efficient information transmission, may provide an intriguing perspective in the present context of a rather high-level view of brain dynamics.
We believe that that the observed scaling may be used as a nontrivial constraint on potential models of large-scale brain behaviour.

The statistical physics interpretation of the Zipf law in terms of an exponential dependence of the density of states on energy~\eqref{e.hagedorn}, provides a very surprising link to the concept of Hagedorn temperature, which appears in various contexts in high-energy physics. The unexpected relevance of this concept to normal brain function is an intriguing result of our work.

In this respect, the relation of the Hagedorn temperature extracted from the density of states to the reference/operating temperature of the effective Boltzmann distribution is of particular interest.
We uncovered a power law scaling with the granularity of the parcellation~\eqref{e.powerlaw}, which shows that the Hagedorn temperature approaches the operating temperature for finer and finer parcellations. 
In the Zipf law language this means that the power $s$ in~\eqref{e.zipf} approaches the classical value $s=1$.
The equality $T_H \sim T_{ref}$ means that for a range of energies the probability distribution of the energies is flat~\eqref{e.measure}.
Going back to the probabilistic interpretation of Eq.~\eqref{e.energy}, this means that even for a large system, states with a wide range of probabilities nevertheless still occur, endowing the system with a very \emph{diverse} dynamical repertoire, a fundamental property which in neurobiology is interpreted as the required microscopic basis of a rich (macroscopic) cognitive behaviour. This is marked contrast to chaotic or to ordered behaviour, where the range of probabilities occuring in the data would be definitely smaller.

Summarizing, the analysis of large scale fMRI brain recordings via a novel machine learning approach  shows the existence of a robust region of a linear $S(E)$ regime, whose range and proximity to the ``operating temperature''  $T_{ref}=1$ increases with the dimensionality of the sampling $N$, i.e. going to finer and finer length-scales. This leads to the finding that the brain operates asymptotically at the Hagedorn temperature.  
The appearance of the linear $S(E)$ dependence demonstrates that a wide range of brain states follow Zipf's law.
The approach presented in this paper on one side allowed the first empirical demonstration of the wide linear regime of a very diverse (critical) density of brain states -- fundamental for normal brain function -- and on the other offers a powerful tool to study other complex systems.

\medskip

\noindent {\textit {Acknowledgments.--}}
This work was conducted under the auspices of the Jagiellonian University-UNSAM Cooperation Agreement and supported by Grant No. 1U19NS107464-01 from the NIH BRAIN Initiative, by CONICET (Argentina) and Escuela de Ciencia y Tecnología, UNSAM, (Argentina) and by the Foundation for Polish Science (FNP) project TEAMNET “Bio-inspired Artificial Neural Networks” (POIR.04.04.00-00-14DE/18-00). 
RJ was also supported by a  Priority Research Area DigiWorld grant under the Strategic Programme Excellence Initiative at the Jagiellonian University (Kraków, Poland).

\clearpage
\appendix

\onecolumngrid

\section{Supplementary information}

\renewcommand{\thefigure}{S\arabic{figure}}
\setcounter{figure}{0}

We include here additional information concerning technical details of the fMRI data pre-processing, checks for the robustness of the results and additional observations.

\medskip
\noindent {\bf fMRI data.} We used the ICA-FIX cleaned resting-state fMRI data for 100 unrelated subjects from the Human Connectome Project \cite{HCPack} Young Adult dataset dowloaded from \url{https://www.humanconnectome.org/study/hcp-young-adult}. 
This includes 4 sessions of approximately 15 min duration each, collected with TR=0.72s. The resulting data were subsequently detrended, band-pass filtered in the range 0.01-0.25Hz and normalized to zero mean and unit standard deviations (using \texttt{nilearn.signal.clean}). 
The first and last 30 frames were dropped, resulting in 1140 frames for each of the 400 sessions. In this way we obtain timeseries of 456000 datapoints length.
Binarization was performed for each session individually, i.e. always using the medians of the specific session  as thresholds.

\begin{figure}[htb]
\includegraphics[width=0.32\textwidth]{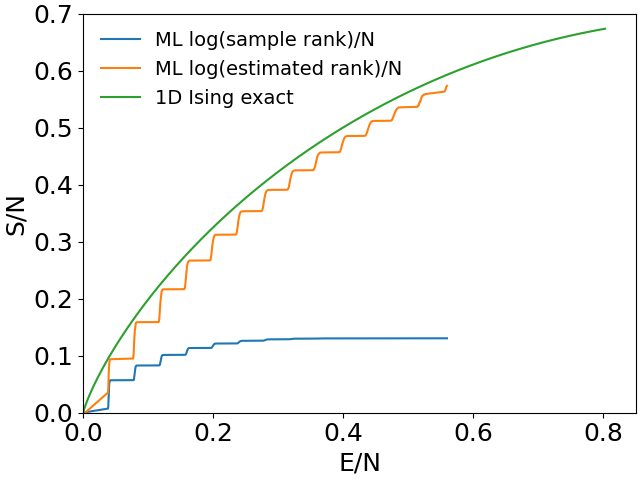}
\includegraphics[width=0.32\textwidth]{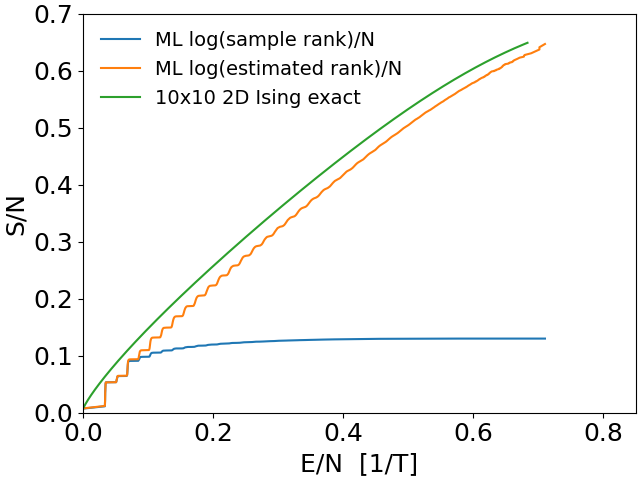}
\includegraphics[width=0.32\textwidth]{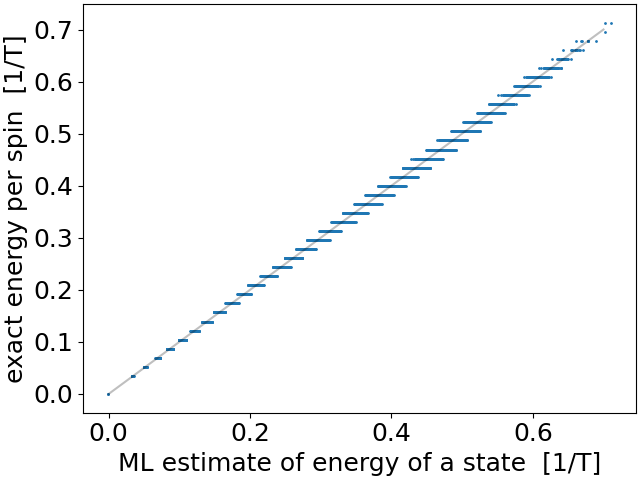}
\caption{Further tests of the machine learning method for systems with $N=100$ spins: noncritical 1D Ising model at $T=1$ (left), critical 2D Ising model at $T=2.3$ (center), comparison of the energies of states extracted from the Monte Carlo data of the 2D Ising model by the machine learning algorithm with the true energies (right). 
\label{fig.moreising}}
\end{figure}

\medskip
\noindent {\bf Further Ising model tests.}
In order to test our method of estimating entropy and energy from a provided set of spin configurations, we performed Monte Carlo simulations using the Metropolis algorithm of the 1D Ising model with $N=100$ and $N=400$ spins at $T=1$ and the 2D Ising model on $10\times 10$ and $20 \times 20$ lattices. 
In order to have comparable overall statistics to the HCP brain fMRI data, in each case we generated 400 sessions of 1140 samples, starting each session from a different random initialization. We used 1000 and 4000 thermalization steps between the MC samples for the models with 100 and 400 spins respectively. 
In Fig.~\ref{fig.moreising}, we show results for the 1D and 2D models with $N=100$ spins. In addition, we directly compare the energies of all individual states present in the MC data to the values predicted by our machine learning algorithm using the \texttt{xgboost} classifier.

\begin{figure}[htb]
\includegraphics[width=0.32\textwidth]{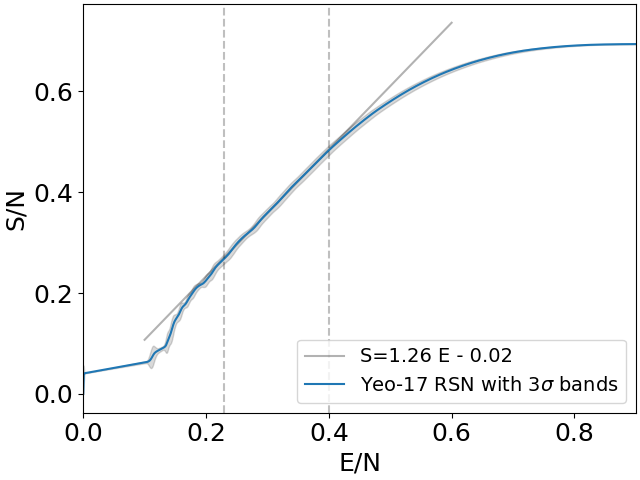}
\includegraphics[width=0.32\textwidth]{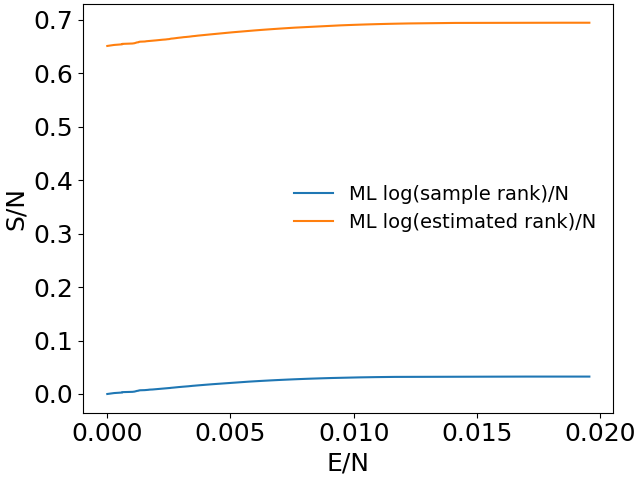}
\includegraphics[width=0.32\textwidth]{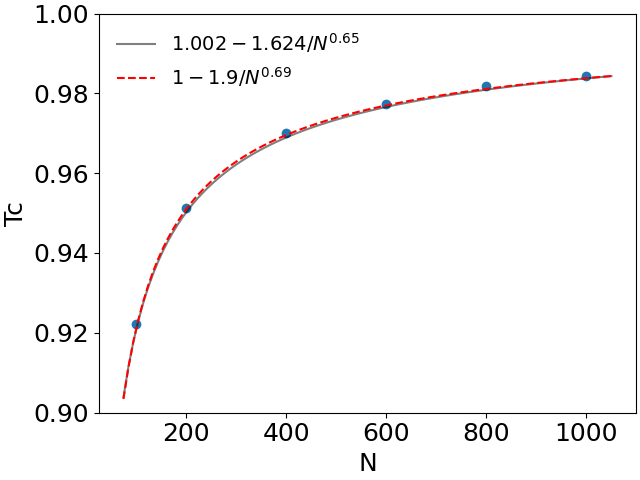}
\caption{$3\sigma$ error bands for the entropy-energy plot for Yeo-17 RSN from 1000 bootstrap datasets (left). $N=400$ parcellation data with each ROI's signal randomly shifted, thus decorrelating the data. Note the very narrow range of energies (probabilities) on the horizontal axis. The predicted probabilities for all states are in the vicinity of $1/2^{400}$ (center). 
Extraction of the critical (Hagedorn) temperature as a function of $N$ from fMRI data with no low-pass filtering. (right). 
\label{fig.randomized}}
\end{figure}

\pagebreak

\noindent {\bf Bootstrap, randomization and robustness tests.} For the Yeo-17 RSN, we performed a bootstrap estimation of the statistical errors. Bootstrap was performed on the session level, i.e. 
we generated 1000 bootstrap datasets by repeatedly randomly selecting 400 sessions with replacement. The entropy-energy curve was obtained by training the \texttt{xgboost} classifiers on each whole bootstrap dataset and subsequently predicting the energies of all $2^{17}$ states in the state-space.
The results are shown in Fig.~\ref{fig.randomized}(left). The entropy-energy curve appears remarkably robust. The determination of the slope is $1.2559 \pm 0.0082$, which translates to the critical temperature $0.7963 \pm 0.0052$.

In order to compare the obtained results with the case of a lack of internal structure of the states, we decorrelated the signals of all ROIs (peformed here for the $N=400$ Schaefer parcellation) by concatenating the sessions, randomly cyclically shifting the signal of each ROI and then again splitting into sessions.
The resulting entropy-energy curve is shown in Fig.~\ref{fig.randomized}(center). The range of energies (probabilities) is very narrow and is concentrated around $1/2^{400}$ (this value is not seen on the plot as we use the convention of setting the energy of the lowest observed state to zero). Such a behaviour is clearly very different from any brain data we studied.

In order to test the influence of filtering of the fMRI data on our results, we repeated the computations leading to asymptotic scaling of the critical temperature with the number of ROIs without performing any low-pass filtering. The result is shown in Fig.~\ref{fig.randomized}(right), which is an analog of Fig.4B in the main text. We observe that the conclusion of the brain operating asymptotically at the Hagedorn temperature still holds with a high accuracy. The scaling exponent is 0.69 (in comparison with 0.74 there) -- both are consistent with a rough value of 0.7. Indeed, comparing the two fits for each set of data, we see that the scaling exponent is reliable to the leading decimal digit.

\medskip
\noindent {\bf Phenomenological Renormalization Group (PRG).} The PRG construction operates on the initial \emph{non-binarized} data. In the present paper, we start with the fMRI data parcellated in the N=1000 Schaefer parcellation, which for the HCP data has actually 998 nonempty ROIs. We compute the $998\times 998$ correlation matrix, identify the pair $i,j$ of ROIs with maximal correlation and form a new signal
\eq
y_1(t) = \f{1}{2}\left(x_i(t) + x_j(t) \right)
\eqx
We then repeat the process with the remaining ROIs, constructing further $y_k(t)$ until we exhaust all ROIs or a single one remains. In the latter case we drop it.
Repeating the above procedure iteratively, we obtain PRG datasets of dimensionality N=499, 249, 124 and 62. For further analysis, we binarize them exactly as the original parcellated data.

\begin{figure}[htb]
    \centering
    \includegraphics[width=0.33\textwidth]{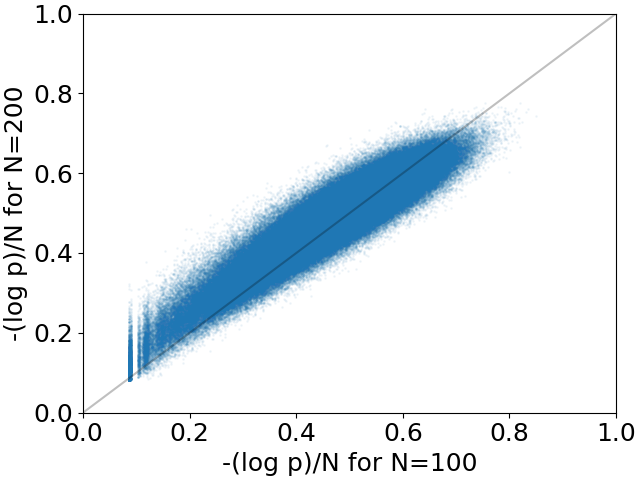}\hfill
    \includegraphics[width=0.33\textwidth]{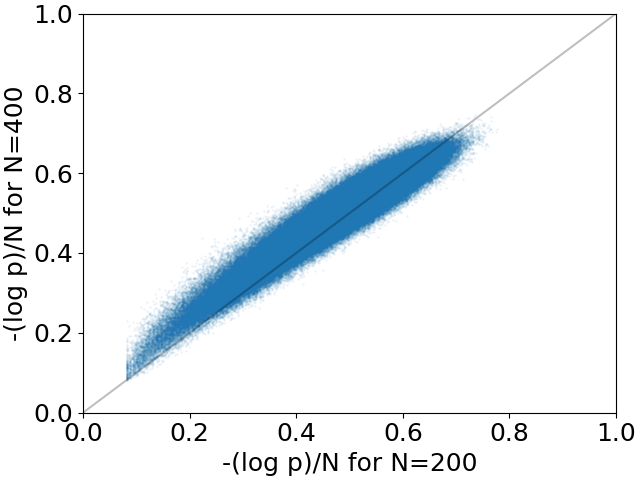}\hfill
    \includegraphics[width=0.33\textwidth]{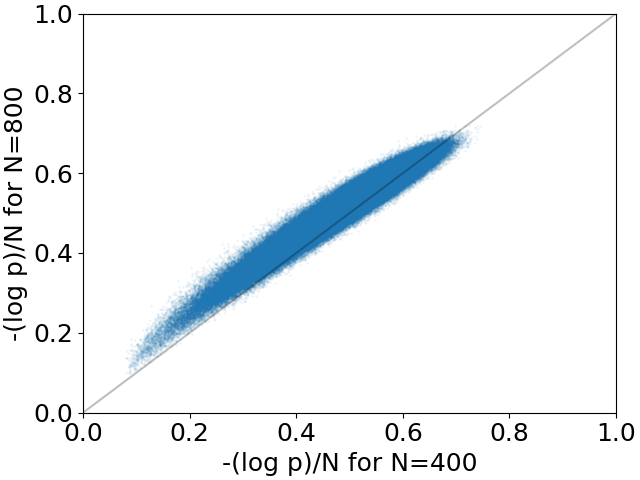}
    \caption{Comparison of the predicted probabilities (energies) for the same states using Schaefer parcellations with $N$ differing by factors of two. The slight systematic deviation of the point clouds from a unit slope suggests that the effective operating temperatures at the corresponding $N$s are sightly different.}
    \label{fig.logpscaling}
\end{figure}

\begin{figure}[htb]
    \centering
    \includegraphics[width=0.33\textwidth]{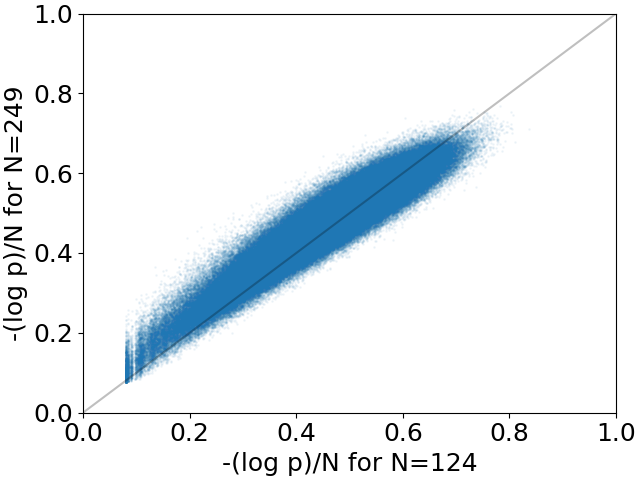}\hfill
    \includegraphics[width=0.33\textwidth]{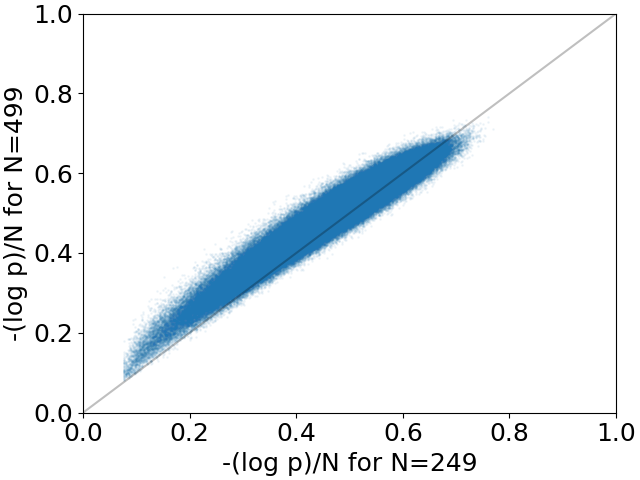}\hfill
    \includegraphics[width=0.33\textwidth]{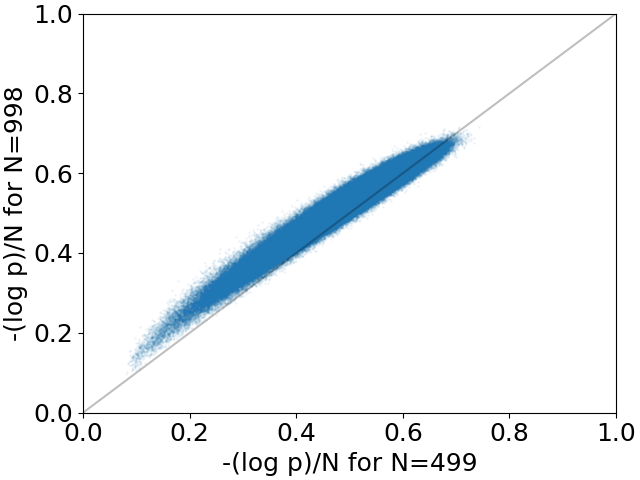}
    \caption{Comparison of the predicted probabilities (energies) for the same states using consecutive steps of the Phenomenological Renormalization Group (PRG). 
    }
    \label{fig.prglogpscaling}
\end{figure}

\medskip
\noindent {\bf Data scale-invariance.}
Let us comment on the relation of the extracted state energies associated with the fMRI data coarse-grained with different granularity. In Fig.~\ref{fig.logpscaling}, we compare the predicted energies for the individual fMRI timesteps between Schaefer parcellations of sizes differring by a factor of two.
We observe a consistency of the predicted energies (defined here as $-\log p$ without any subtraction) normalized to the size of the system.
In Fig.~\ref{fig.prglogpscaling}, we observe a similar behaviour when applying the Phenomenological Renormalization Group construction to the $N=1000$ Schaefer parcellation.
These results can be interpreted as a signature of approximate scale invariance of the system, as we can think of this computation as a comparison of hamiltonians in various coarse-grainings in the Renormalization Group sense.

\pagebreak

\noindent {\bf Details on the machine learning method.} 
In all applications of the machine learning method we use the \texttt{xgboost} classifier~\cite{xgboostbis}, with the settings from~\cite{ENTROPYMLbis}, which were optimized for a $20\times 20$ 2D Ising model, i.e. \texttt{max\_depth=3} and \texttt{n\_estimators=50}.
The computations were done on a GPU employing \texttt{tree\_method='hist'} for efficiency.
As the code~\cite{entropycode} released with ref.~\cite{ENTROPYMLbis} was dedicated just for computing the overall entropy of the data, we provide below the pseudo-code for computing the logarithms of probabilities of the individual samples from the data.
In principle one can use here any machine learning classifier which returns probabilities of the predicted class.
The quality of the classifier or its parameters can be assessed by evaluating the Shannon entropy from the obtained probabilities and choosing the classifier or its settings which yield the lowest entropy.

\begin{algorithm}[H]
  \caption{Estimate the probabilities of individual samples in the provided data}
  \label{alg.rdag}
\begin{algorithmic}[1]
  \STATE {\bfseries Input:} binary data $X^{all}$ of size $S \times T \times N$,\\ 
  \hspace{1.08cm} number of sessions $S$,\\
  \hspace{1.08cm} number of timesteps in a session $T$,\\ 
  \hspace{1.08cm} dimensionality $N$\\
  \STATE {\bfseries Output:} logarithms of probabilities of all states: $logp$ of size $S \times T$
  \STATE extract the first spin $y_{s,t} \equiv X^{all}_{s,t,1}$\\
  \STATE compute probability of the first spin $p_1 = mean(y)$\\
  \STATE initialize $logp_{s,t} = y_{s,t} \log p_1 + (1-y_{s,t})\log(1-p_1)$\\
  \FOR{$i=2$ {\bfseries to} $N$}
  \STATE construct training data for predicting spin $i$: $X = X^{all}_{:,:,1..i-1}$
  \STATE construct target for predicting spin $i$: $y = X^{all}_{:,:,i}$
  \STATE randomly split the sessions into 5 training and testing cross-validation (CV) folds
  \FOR{$X^{train}, y^{train}, X^{test}, y^{test}$ in each CV fold}
  \STATE train a ML classifier $clf$ to predict $y^{train}$ from $X^{train}$
  \STATE predict probabilities for the samples in $test$: $p^{test}=clf(X^{test})$
  \STATE update $logp$ in the $test$ subset of the data:
  \[logp^{test} \to logp^{test} + y^{test}\log p^{test} +(1-y^{test}) \log(1-p^{test})\]
  \ENDFOR
  \ENDFOR
\end{algorithmic}
\end{algorithm}

For performing the analysis of the present paper, we need the probabilities of \emph{distinct} states $\alpha$ in the data, thus extracting the subset of the results $p_\alpha$ returned by the algorithm corresponding to these distinct states. 
Moreover, in order to estimate the multiplicities in Eq. (6) in the main text, we need also the occurrence counts of these distinct states to get
\eq
p_\alpha^{empirical} = \frac{\text{occurrence count of state $\alpha$ in data}}{\text{total number of samples}}
\eqx
Consequently, we may estimate the Shannon entropy of the data through
\eq
S \equiv -\left\langle \log p\right\rangle = -\sum_{distinct\; \al} p^{empirical}_\al \log p_\al = -\f{1}{ST} \sum_{s,t}  \log p_{s,t}
\eqx

\bigskip

The github repository \url{https://github.com/rmldj/brain_hagedorn_paper} contains the code and the binarized fMRI data to reproduce the examples in Figure 1 of the paper.

\end{document}